\documentclass[draftcls,onecolumn, 12 pt]{IEEEtran}
\ifCLASSINFOpdf
\else
\fi

\usepackage{graphicx}
\begin{document}
%
% paper title
% can use linebreaks \\ within to get better formatting as desired
\title{Morphological Detector\\ for Multilevel Signals in $\epsilon$ - Noise
}

% author names and affiliations
% use a multiple column layout for up to three different
% affiliations
\author{\IEEEauthorblockN{Sandy Stepanov, Anastasios Venetsanopoulos}
\IEEEauthorblockA{\\Ryerson University\\
Toronto, Canada\\
Email: sandy@ee.ryerson.ca}}

%\and
%\IEEEauthorblockN{Homer Simpson}
%\IEEEauthorblockA{Twentieth Century Fox\\
%Springfield, USA\\
%Email: homer@thesimpsons.com}
%\and
%\IEEEauthorblockN{James Kirk\\ and Montgomery Scott}
%\IEEEauthorblockA{Starfleet Academy\\
%San Francisco, California 96678-2391\\
%Telephone: (800) 555--1212\\
%Fax: (888) 555--1212}}

% conference papers do not typically use \thanks and this command
% is locked out in conference mode. If really needed, such as for
% the acknowledgment of grants, issue a \IEEEoverridecommandlockouts
% after \documentclass

% for over three affiliations, or if they all won't fit within the width
% of the page, use this alternative format:
%
%\author{\IEEEauthorblockN{Michael Shell\IEEEauthorrefmark{1},
%Homer Simpson\IEEEauthorrefmark{2},
%James Kirk\IEEEauthorrefmark{3},
%Montgomery Scott\IEEEauthorrefmark{3} and
%Eldon Tyrell\IEEEauthorrefmark{4}}
%\IEEEauthorblockA{\IEEEauthorrefmark{1}School of Electrical and Computer Engineering\\
%Georgia Institute of Technology,
%Atlanta, Georgia 30332--0250\\ Email: see http://www.michaelshell.org/contact.html}
%\IEEEauthorblockA{\IEEEauthorrefmark{2}Twentieth Century Fox, Springfield, USA\\
%Email: homer@thesimpsons.com}
%\IEEEauthorblockA{\IEEEauthorrefmark{3}Starfleet Academy, San Francisco, California 96678-2391\\
%Telephone: (800) 555--1212, Fax: (888) 555--1212}
%\IEEEauthorblockA{\IEEEauthorrefmark{4}Tyrell Inc., 123 Replicant Street, Los Angeles, California 90210--4321}}

% use for special paper notices
%\IEEEspecialpapernotice{(Invited Paper)}

% make the title area
\maketitle

\begin{abstract}
\boldmath
% \boldmath inogda ispolzyiyt non bold
The novel approach was developed for multilevel signal detection in channels with impulsive nonGaussian noise. This approach consists of using morphological nonlinear image filtration principles for two dimensional signals. It is a new method of signal demodulation, using three - dimensional image processing algorithms. Successful results of this morphologic detector encourage more investigation towards using image processing theory and algorithms for two dimensional signal processing. As can be seen in the example in section IV, this new approach of reusing well developed and extensively developing image processing has significantly improved performance.
\end{abstract}
% IEEEtran.cls defaults to using nonbold math in the Abstract.
% This preserves the distinction between vectors and scalars. However,
% if the conference you are submitting to favors bold math in the abstract,
% then you can use LaTeX's standard command \boldmath at the very start
% of the abstract to achieve this. Many IEEE journals/conferences frown on
% math in the abstract anyway.

% no keywords

% For peer review papers, you can put extra information on the cover
% page as needed:
% \ifCLASSOPTIONpeerreview
% \begin{center} \bfseries EDICS Category: 3-BBND \end{center}
% \fi
%
% For peerreview papers, this IEEEtran command inserts a page break and
% creates the second title. It will be ignored for other modes.
\IEEEpeerreviewmaketitle

\section{Introduction}
%% no \IEEEPARstart
%This demo file is intended to serve as a ``starter file''
%for IEEE conference papers produced under \LaTeX\ using
%IEEEtran.cls version 1.7 and later.
%% You must have at least 2 lines in the paragraph with the drop letter
%% (should never be an issue)
%I wish you the best of success.
 The widely used Gaussian approach for noise pdf relies on assuming that by  using limiters the impulsive noise can be transformed to approximately a Gaussian one. This approach works for the two-level case, but for multilevel signals  this approach should be reviewed. Consequently, it is not surprising that literature about the comparison between actually achieved and expected communication system performance is very rare, since the expected difference has a dramatic drop in performance. Reference \cite{YooShin_QAM} highlights a remarkable analytical example of performance deterioration due to impulsive nongaussianity.
Known practical ways to achieve the impulse noise immunity are: empirical blanking or cutting noise realization by limiters, using nonparametric statistics theory, and detectors based on Robust Statistic Theory.  The suggested approach belongs to the nonparametric statistical theory  branch, and specifically to nonlinear filtering. The essential methodological principle is to use three-dimensional image processing for signals conventionally described by a two dimensional model. It is advantageous to use  a large bank of image processing algorithms in the decoder design. We show here that a tremendous improvement of performance can be achieved even if we are attempting a proof of principle for the first time. Further investigations  in this direction are expected to bring more performance gain by a variety of image processing approaches. Our intention of opening a new way for demodulation design is successful and promising as it can be seen from achieved morphological detector performance. The realization complexity is moderate, since all morphological filter calculations are accomplished for logical operations for numbers "1" and "0".
Proof of principle  is done by using a two-level signal, when it is assumed that the limiter around these signal levels can not be used, since other signal levels can also be expected. Such simplicity is convenient for detecting observing and analyzing process features, when general game rules for conventional multilevel signal demodulation are adhered to:
       the signal and noise should pass through a receiver filter; after a nonliner signal  is detected by a conventional matches filter detector. But the signal can be at a wide range (much more than an unnoisy signal range).

Morphological filtering (MoF) is used as specific signal transformation to cut impulsive realizations from the signal. As can be seen from the performance curves below, as the impulsivity of noise increases the MoF signal BER approaches the ideal one, when an optimal MAP detector is placed at receiver input and the full statistical signal description is known. The result is: significant a high level of efficiency is achieved, despite unknown noise parameters. In other words, an efficient invariant to noise impalsivity detection is attained. It is remarkable that the performance margin is minor for only Gaussian noise. If necessary, the switcher between impulsive and nonimpulsive channel environments can be used to eliminate this shortage for  practical use.
Last but not least, the gain from morphological detector usage is increased significantly when the error correction coding is used, since coding BER is dramatically expanded even with minor BER improvement at the  detection level. What is more, the gain from shifting from one area where the code is inefficient to an area where the code is efficient can not be expressed by the BER improvement measure, since the improvement is qualitative rather than quantitative. In particular this improvement  is achieved by using the suggested signal processing.

\section{System Model}
The mixed Gaussian noise ($\epsilon$ - noise) model is a simple approach \cite{Aazhang_Poor}, \cite{Kassam_book}   to the Middleton general Gaussian noise description \cite{Middleton}. It is shown analytically \cite{Miller} that this noise model is a realistic choice. Analytically, $\epsilon$ - noise pdf for noise realization $\xi$, at receiver antenna input is described by
\begin{equation}\label{eq:1p1}
\rho(\xi)=(1-\varepsilon)\varphi(\xi,0,\sigma_1)+
\varepsilon\varphi(\xi,0,\sigma_2),
\end{equation}
where\\
 $\rho(\xi)$ is the pdf of the stochastic process $\xi$;\\
$\varphi(\xi,0,\sigma_1)$ is the gaussian pdf of the background noise   stochastic process   with mean 0 and  std
$\sigma_1$;
\newline $\varepsilon$ is the probability of the impulsive noise; relatively, (1-$\epsilon$) is the probability of background noise only;
\newline
$\varphi(\xi,0,\sigma_2)$ is the gaussian pdf of the stochastic process of the sum of background and impulse noise with mean  0  and std  $\sigma_2
\gg \sigma_1$   .

Physically, this model has a clear interpretation: with probability (1- $\epsilon$) the background noise is generated by a noise source; with probability $\epsilon$, impulse noise and background noise are generated. Both pdfs are Gaussian; therefore, the overall pdf for the sum of background and impulse noise is Gaussian, too.
	In the case of impulse noise generation by military jamming  \cite{Holmes_book}, the description is similar.
	The linear filtration in receiver input is a realistic scenario, so detector input noise is the convolution of noise with impulse response h(t) of the receiver filter Matlab was used for h(t) calculation, so $\sum\limits_{i} h_i =1$.
	The effective noise rectification by a limiter can be used for only two signal levels, whereas modern communication systems use multilevel signals. WiFi and WiMax standards are examples of such systems. It is natural to look for  algorithms with limiter abilities applicable to multilevel signals, too. Where "whitening" relates to the whitening operation of color noise, the  term "gaussining" applies to the approach taken in this paper.
	The filtering process makes the $\epsilon$-noise vary gradually according to h(t) length. This effect  means that now we have "mountains"  and "valleys" with durations approximating h(t) length. For example,  we used h(t) with a number of significant samples, amounting to nearly $10$ percent of the  number of  symbol samples (see Fig.1). In this case, we used $70$ samples.
\begin{figure}[!htp]
\centerline{\includegraphics[width=3in]{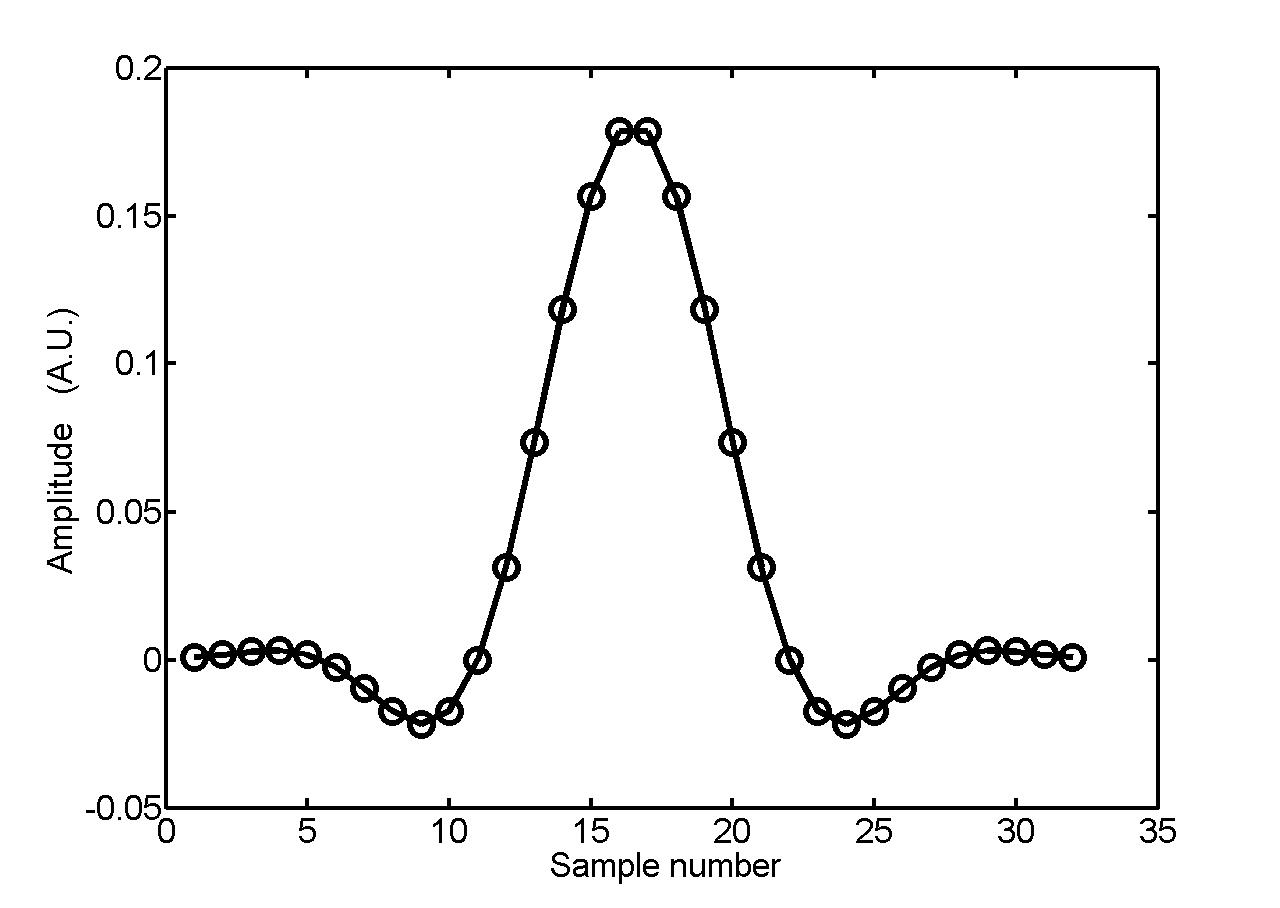}} \caption{ The input filter impulse response
 } \label{Fig_2}
\end{figure}
%\begin{figure*}[htp]
%\centerline{\includegraphics[width=2in]{fig_2_transm_1}} \caption{ The qq
% } \label{Fig_2}
%\end{figure*}

%Fig_2_transmitted_1

 %{\includegraphics[width=3in]{Fig_2_transmitted_1}}

%\begin{figure}
%  % Requires \usepackage{graphicx}
%  \includegraphics[width=4in]{Fig_2_transmitted_1}\\
%  \caption{rr}\label{tt}
%\end{figure}

%\begin{figure}
%  % Requires \usepackage{graphicx}
%  \includegraphics[width=4in]{pic3p1}\\
%  \caption{rr}\label{tt}
%\end{figure}

%\begin{figure}[!t]
%\centering
%\includegraphics[width=2.5in]{fig_2_transm_1}
%\caption{Detection example}
%\label{fig_2}
%\end{figure}

	The critical issue of the model is to find the potential bound. This bound should be used for algorithm effectiveness estimation and for trade-off design in practical use, when a designer optimizes the algorithm by balancing effectiveness of mutual criteria against cost.  The first straight-forward decision is to use Maximum a Posteriori (MAP) detector for pdf (1) when noise is not distorted by a filter. This results in a  nonfeasible detector at antenna input.
	When the parameters $\epsilon$, $\sigma_1$, $\sigma_2$ are known, the detector determines that the transmitted symbol is more probable if

\begin{equation}\label{eq_2}
    \prod_i \rho(\xi_i) > \prod_i \rho(\xi_i + e_i)
\end{equation}

where
  $e_i$ is the difference between  sample number $i$ of the transmitted symbol and its alternative.

	The BER curve for this detector seems to be good enough to be potentially achieved.
	The parameter $\epsilon$ is difficult to estimate, so the next candidate for potential reference is the detector, which some how knows whether the background noise or the sum of background noise and impulsive noise is at an input of the antenna. The decision is made in favor of the transmitted symbol if
%\begin{equation}\label{eq_3}
%    \prod_i  (k_i  (1-\varepsilon)\varphi(\xi,0,\sigma_1)+ (1-k_i)
%\varepsilon\psi(\xi,0,\sigma_2) ) >   \prod_i (k_i  (1-\varepsilon)\varphi(\xi+e_i,0,\sigma_1)+ (1-k_i)
%\varepsilon\psi(\xi+e_i,0,\sigma_2) )
%\end{equation}

%\begin{eqnarray}\label{eq_3}
%&\textbf{}{\prod_i}  (k_i  (1-\varepsilon)\varphi(\xi,0,\sigma_1)+ (1-k_i)\varepsilon\varphi(\xi,0,\sigma_2) ) >\nonumber\\
%& \prod_i (k_i  (1-\varepsilon)\varphi(\xi+e_i,0,\sigma_1)+ (1-k_i)\varepsilon\varphi(\xi+e_i,0,\sigma_2) )
%\end{eqnarray}

%\begin{eqnarray}\label{eq_3}
%&\textbf{}{\prod\limits_{i}}  (k_i  (1-\varepsilon)\varphi(\xi,0,\sigma_1)+ (1-k_i)\varepsilon\varphi(\xi,0,\sigma_2) ) >\nonumber\\
%& \prod\limits_{i} (k_i  (1-\varepsilon)\varphi(\xi+e_i,0,\sigma_1)+ (1-k_i)\varepsilon\varphi(\xi+e_i,0,\sigma_2) )
%\end{eqnarray}

\begin{eqnarray}\label{eq_3}
&  \prod\limits_{i}  (k_i  (1-\varepsilon)\varphi(\xi,0,\sigma_1)+ (1-k_i)\varepsilon\varphi(\xi,0,\sigma_2) ) >\nonumber\\
& \prod\limits_{i} (k_i  (1-\varepsilon)\varphi(\xi+e_i,0,\sigma_1)+ (1-k_i)\varepsilon\varphi(\xi+e_i,0,\sigma_2) )
\end{eqnarray}

%\limits_{i=0}

%\begin{eqnarray}
%Z&{}={}&x_1 + x_2 + x_3 + x_4 + x_5 + x_6\nonumber\\
%&&+a + b\\
%&&+{}a + b\\
%&&{}+a + b\\
%&&{+}\:a + b
%\end{eqnarray}

%\begin{equation}\label{eq_33}
%    \prod_i  (k_i  (1-\varepsilon)\varphi(\xi,0,\sigma_1)+ (1-k_i)
%\varepsilon\psi(\xi,0,\sigma_2) ) >
%\end{equation}
%\begin{center}
%    $\prod_i (k_i  (1-\varepsilon)\varphi(\xi+e_i,0,\sigma_1)+ (1-k_i)
%\varepsilon\psi(\xi+e_i,0,\sigma_2) )$
%\end{center}

%\begin{eqnarray}
%% \nonumber to remove numbering (before each equation)
%  ss &=& ff \\
%  yy &=& gg
%\end{eqnarray}
where
	$k_i$ – is the indicator, whether the  noise is only background noise or the sum of background noise and impulsive noise.
	Last but not least, the model is designed for proof  of the principle case, so it should be as general as possible. For  this reason, a rectangular symbol shape was chosen. The test case investigates model behavior for two signal levels. Here, limiting or blanking by using information about possible signal levels is impossible, since it is assumed that other signal levels exist.

\section{Morphological Detector}
As mentioned above, the signal after the input filter looks like a mountain  landscape image, so it is logical to use image processing methods. Moreover, there is impulse noise; therefore, image processing methods for nonGaussian noise can be a good choice \cite{Tas}.  Analysis of \cite{Tas}, leads to the conclusion that MoF can be used for signal smoothing, in order to eliminate distractive impulse noise influence.
	The first view decision is to use a conventional MoF consisting of a sequence of opening and closing operations \cite{Tas}
\begin{equation}\label{eq_4}
    \widetilde{S} = (S \circ SE)\bullet SE,
\end{equation}

where

$S$ is signal image;

$SE$ is structure element, in the current example a line with length 15 is used for $SE$.

However, as shown below, the double use of  MoF is required when the second use is for the symmetric version of the signal image. We need two interfaces: one to convert the data signal $s$ to image $S$ and the second to convert filtered $S$ to a signal, in order to make a decision.
	The first interface starts from

$S(i,j) = 1, i = 1,2,...N$   and $j = 1,2,...M,$

 where $N$ is the number of signal discretization levels and $M$ is the number of samples for symbol. In our example case $N = 300$ and  $M = 70$.
	It is important to stress  that in OFDM systems, values of $M$ tend to be large. Therefore, there is no requirement for dramatic hardware changes to implement MoF in OFDM systems, which use one-carrier signal, too. The next step is

%\begin{equation}\label{eq 5}
%    S[i,j=1:(S_{int}(i)+V/2)]=0
%\end{equation}
%
%
%\begin{center}
%    i=1,2,...M!!!!
%\end{center}

\begin{eqnarray}\label{eq_5}
&  S[i,j=1:(S_{int}(i)+V/2)]=0,
& i=1,2,...M,
\end{eqnarray}

where
 $V = N/2$,
 $S_{int}  = round(s*K)$, i.e. $S_{int}$ is the  quantized  version of the signal, using scale factor \emph{K}.
	
The second interface calculates the filtered signal

\begin{equation}\label{eq 6}
    s_1(i)=[\sum_j \widetilde{S}(i,j)]-V.
\end{equation}

	One of the main problems is the asymmetry of MoF. We overcome this by calculating the symmetrical image $S_s$ as described below, and applying MoF once again to this new image converting the result to signal $s_2$. Mathematically, the symmetrical image $S_s$ is calculated, according to following two steps. The first step is
\begin{center}
  $S_s(i,j) = 1, i = 1,2,...N$   and $j = 1,2,...M$.
\end{center}
The second step is
\begin{equation}\label{eq 7}
    S_s[i,j=1:(-S_{int}(i)+V/2)]=0, i=1,2,...M.
\end{equation}

Then, MoF is applied:
\begin{equation}\label{eq_8}
    \widetilde{S}_s = (S_s \circ SE)\bullet SE.
\end{equation}

The transformation to a signal is carried out with analogy to (6) using as answer the value $s_2$.
The final operation is to find the average signal

\begin{equation}\label{eq 9}
    s_r=(s_1+s_2)/2
\end{equation}

%[ \alpha_k(s) = \frac{\Gamma_k(s)}{\sum_{j=1}^n \Gamma_j(s)}
%   \quad\quad  (k=1,\ldots,n), \]

and to find which level "1" or "-1" is more probable, by using as answer the value

\begin{equation}\label{eq_10}
    R = sign \sum_i s_r(i).
\end{equation}

%(\ref{eq:1p1})

%\cite{feher_impuct}

\section{Simulation Results}
General understanding of MoF detection can be obtained from looking at Fig. 2, showing the detecting process.
\begin{figure}[!htp]
\centerline{\includegraphics[width=3in]{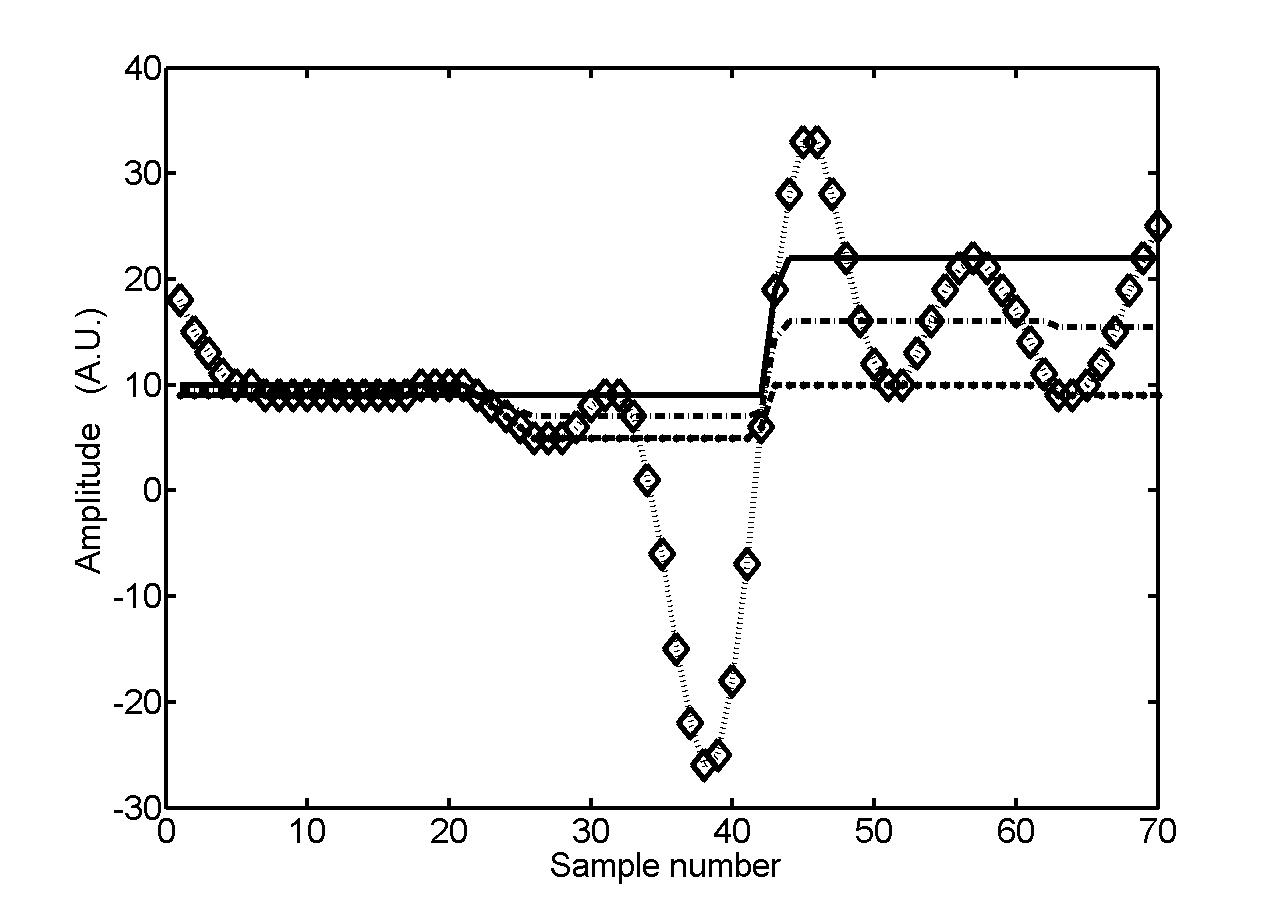}} \caption{ The detecting process illustration for right decision that level $"1*K"$ was transmitted ( $K=10$), solid line corresponds to MoF applied to signal image , dashed line corresponds to MoF applied to symmetrical  signal signal iamge, the dashed line with dots corresponds to average decision.
 } \label{Fig_2}
\end{figure}
From Fig. 2, it can be seen how strong noise pulses in the signal can be cut by the MoF. Each of these pulses can shift the decision to the wrong polarity, but all of them are easily removed by MoF. In spite of the minor negative pulse which is not aligned, the decision that the overall signal is negative is right. The additional resource - to make algorithm enable side pulses to be cut -  can be simply  applied to the waveform in Fig. 2. For example the signal can be synthetically prolonged  to the left and right, but this is beyond the scope of this paper, since we only show the proof of principle for MoF detection.
From a theoretical point of view, it is important to investigate the tendency of BER change  variations in (a) impulse noise probability; (b) Gaussian noise level; (c) impulse noise level.
	The relations between noise and signal levels may be conveniently investigated when BER curves are depicted for noise std values, rather than for the relation between powers of noise and signal. The reason for this is that it is important for the morphological filter to be able  to cut the strong noise pulses. Therefore, the  BER curves are represented by BER dependency on the std of the sum of the impulse noise and background noise for a fixed background noise level in contrast to the more conventional method of BER performance representation. On the one hand, the impulse noise level  must be large enough to be able to cause errors. Therefore, it is much greater than the worst background noise level from Fig. 3, where only Gaussian noise is applied.
\begin{figure}[!htp]
\centerline{\includegraphics[width=3in]{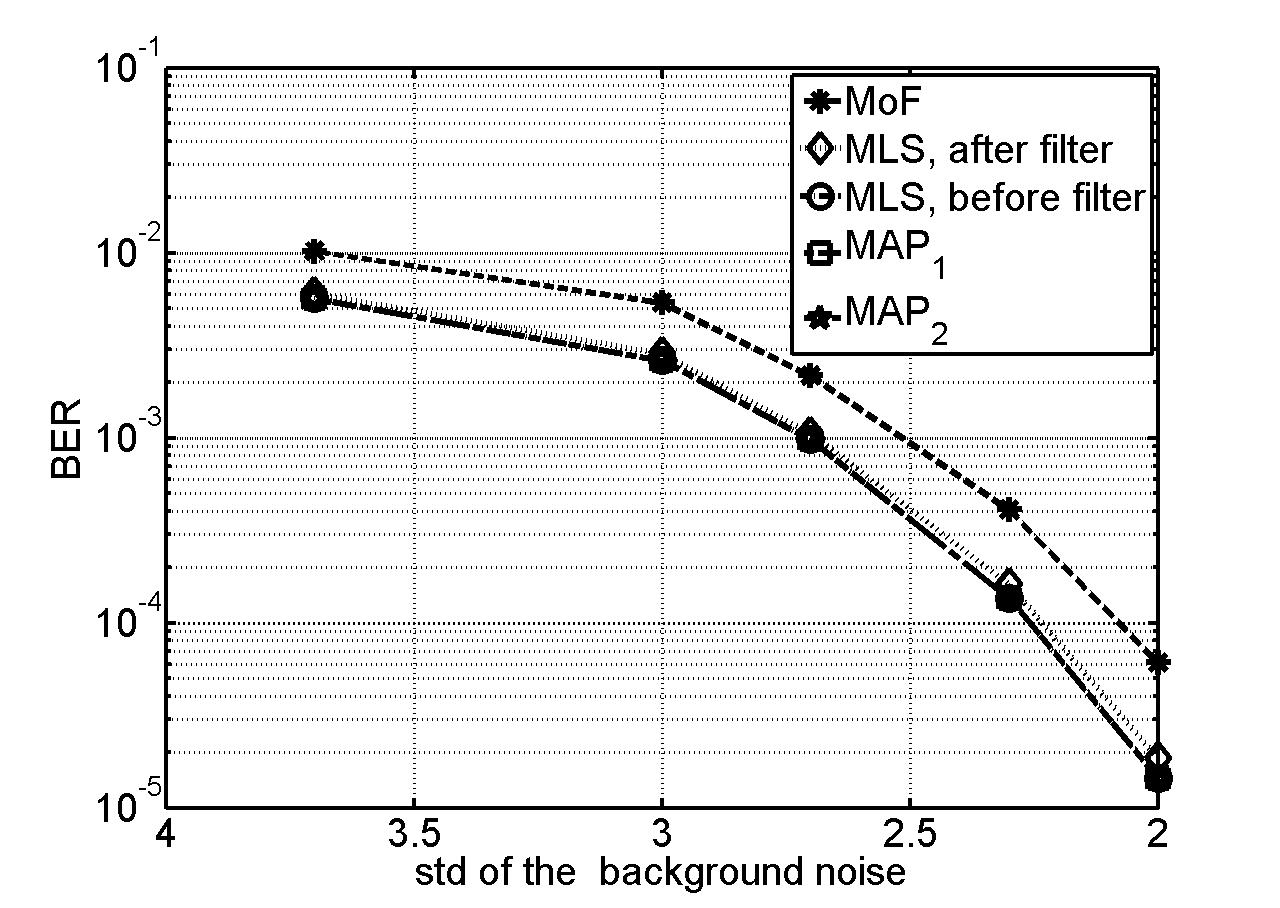}} \caption{ Performance in only Gaussian noise,$MAP_1$ stands for MAP  detection according (2), $MAP_2$ stands for MAP  detection according (3).
 } \label{Fig_3}
\end{figure}
On the other hand, the impulse noise level significantly but not greatly exceeds 256 QAM signal maximum levels. The analysis of Fig. 3 shows modest a margin for the MoF detector for only Gaussian noise conditions. For nonGaussian noise as the support point, the set $\epsilon=0.01$ and $\sigma_1=2$ is chosen and then: (a) impulse noise probability was reduced to $\epsilon=0.001$; (b) the background noise level was reduced to $\sigma_1=1$. The initial impulse noise set was represented by a typical value of impulse noise probability, when the background  noise role is to provide reasonable  BER for practical use (see Fig. 4).
\begin{figure}[!htp]
\centerline{\includegraphics[width=3in]{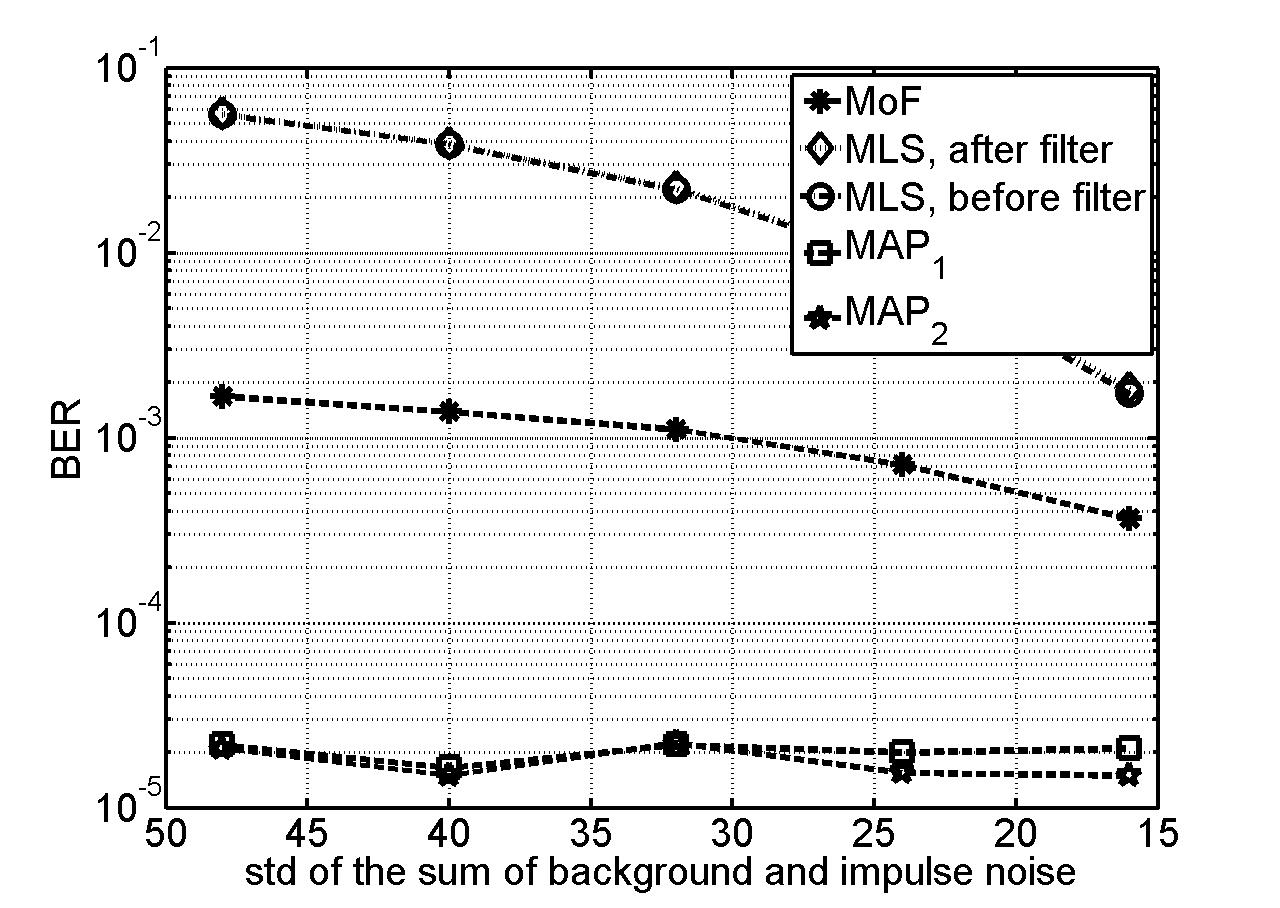}} \caption{ Performance for $\varepsilon$ noise for $\epsilon=0.01$ and $\sigma_1 = 2$, $MAP_1$ stands for MAP  detection according (2), $MAP_2$ stands for MAP  detection according (3).
 } \label{Fig_4}
\end{figure}
	The gain from using MoF is obvious. The next  point is to check how noise impulsiveness, increased by reduction of epsilon, influences performance. It is possible to see from Fig. 5, that MoF gives further BER closure to optimal detectors and its behavior is the same as the optimal one, when $\epsilon$ is reduced by a factor of 10.
\begin{figure}[!htp]
\centerline{\includegraphics[width=3in]{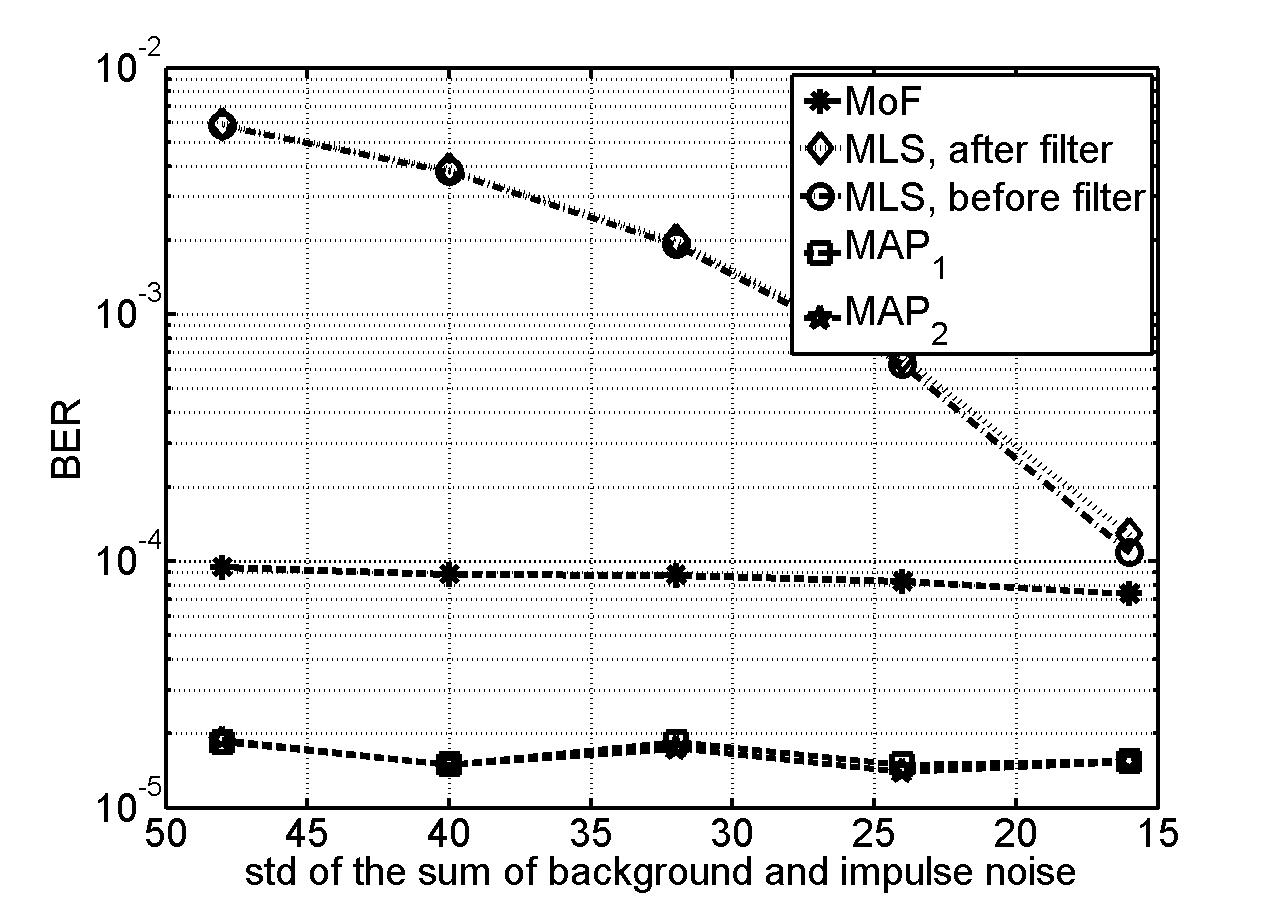}} \caption{ Performance for $\varepsilon$ noise for $\epsilon=0.001$ and $\sigma_1 = 2$ , $MAP_1$ stands for MAP  detection according (2), $MAP_2$ stands for MAP  detection according (3).
 } \label{Fig_5}
\end{figure}
The  BER for MoF is limited by MoF performance in only Gaussian noise. In Fig. 5  MoF BER is approximately $10^{-4}$ and in Fig. 3 there is the same BER for the same $\sigma_1=2$. The behavior of MoF is similar to that of an optimal detector. Optimal detection performance at Fig. 5 is defined by BER in only Gaussian noise for the same $\sigma_1=2$. Both optimal detectors have the same performance, which is defined only by background noise. This shows that an efficient detector can not be given parameter $\epsilon$, but only should recognize whether the noise generated results from background noise or impulse noise. This once again shows that empirical limiting has potentially near optimal detector abilities, since the use of the limiter enables the detector to make a decision only by analyzing low level background noise. One important conclusion follows from this observation: influence of impulse noise can be potentially eliminated, so further improvement of MoF is possible and will bring significant gain. Further noise impulsivity increased by background noise influences reduction, leading to further  detection efficiency of MoF, as can be seen in Fig. 6.
\begin{figure}[!htp]
\centerline{\includegraphics[width=3in]{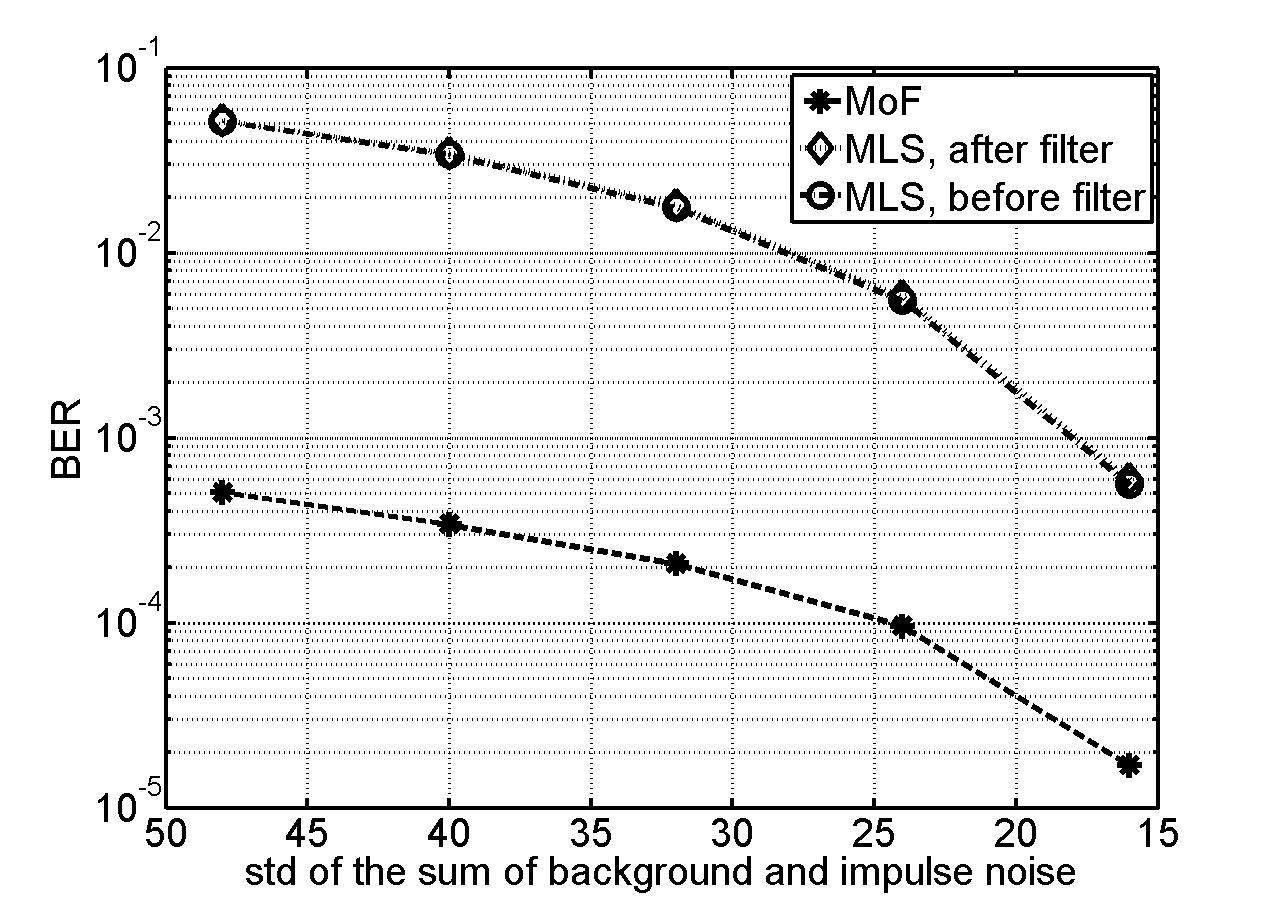}} \caption{ Performance for $\varepsilon$ noise for $\epsilon=0.01$ and $\sigma_1 = 1$.
 } \label{Fig_6}
\end{figure}
As a result, all mistakes are caused by impulse noise, and MoF is efficient particularly for this kind of noise. The optimal detector  performance can not be received by simulation, since background noise influences is negligible for the condition of  Fig. 6.  The demonstration of quantitative gain is very impressive. Where as MoF is not given any information about noise pdf parameters when optimal  detectors are given all the information. One more impressive  note is that MoF accomplishes only binary numbers operations in comparison with the very complicated calculation of optimal detectors.

The simulation  was carried out  until at least $100$ mistake symbols occurred to ensure reliable results.

\section{Conclusion}
A new nonparametric detection approach for using image signal processing for multilevel signal which bears   digital data is suggested. The high potential ability of this approach was shown using a detector based on a Morphological filter. This successful proof of the principle  encourages  more investigation of  image processing usage in the area of conventional detection.

% conference papers do not normally have an appendix

% use section* for acknowledgement

% trigger a \newpage just before the given reference
% number - used to balance the columns on the last page
% adjust value as needed - may need to be readjusted if
% the document is modified later
%\IEEEtriggeratref{8}
% The "triggered" command can be changed if desired:
%\IEEEtriggercmd{\enlargethispage{-5in}}

% references section

% can use a bibliography generated by BibTeX as a .bbl file
% BibTeX documentation can be easily obtained at:
% http://www.ctan.org/tex-archive/biblio/bibtex/contrib/doc/
% The IEEEtran BibTeX style support page is at:
% http://www.michaelshell.org/tex/ieeetran/bibtex/
%\bibliographystyle{IEEEtran}
% argument is your BibTeX string definitions and bibliography database(s)
%\bibliography{IEEEabrv,../bib/paper}
%
% <OR> manually copy in the resultant .bbl file
% set second argument of \begin to the number of references
% (used to reserve space for the reference number labels box)

\end{document}